# Formulating Subgroup Discovery as a Quantum Optimization Problem for Network Security

Samuel Spell
*MU Quantum Innovation Center*
*Department of Electrical Engineering and Computer Science*
*University of Missouri - Columbia*
Columbia, MO, United States
sesxr2@missouri.edu

Chi-Ren Shyu
*MU Quantum Innovation Center*
*Department of Electrical Engineering and Computer Science*
*University of Missouri - Columbia*
Columbia, MO, United States
shyuc@missouri.edu

*Abstract*: **While current network intrusion detection systems achieve satisfactory accuracy, they often lack explainability. Subgroup Discovery (SD) addresses this by building interpretable rules that characterize feature interactions associated with attack traffic. With large datasets, classical heuristic beam search methods struggle with exponentially scaling search spaces and can prune critical multi-feature interactions. This paper introduces a quantum-enhanced pipeline for SD applied to network intrusion detection using NSL-KDD, formulating SD as quantum optimization for the first time. By encoding feature selection as a Quadratic Unconstrained Binary Optimization (QUBO) and solving it via the Quantum Approximate Optimization Algorithm (QAOA) on IBM Quantum hardware (ibm_pittsburgh), the pipeline identifies subgroups of network features that discriminate normal from attack traffic. A least-squares regression QUBO formulation fits the Weighted Relative Accuracy (WRAcc) landscape over feature subsets, with surrogate sampling for larger QUBOs. Results are benchmarked against exhaustive enumeration and Beam Search using ratios for Hamiltonian quality and WRAcc. Hardware scaling experiments on ibm_pittsburgh (10-30 qubits) reveal that QAOA at depth p = 1 shows WRAcc ratios of 0.983 at 10 qubits, 0.971 at 15 qubits, 0.855 at 20 qubits, and 0.624 at 25 qubits, degrading to 0.039 at 30 qubits as circuit noise dominates, establishing an empirical NISQ scaling boundary. Results demonstrate that QAOA discovers subgroups competitive with classical heuristics and finds multi-feature interaction patterns that greedy Beam Search prunes, with QAOA-unique subgroups achieving up to 99.6% test precision. This work establishes a framework for quantum combinatorial optimization in cybersecurity and characterizes hardware scaling for NISQ devices.**

## I. INTRODUCTION

As network infrastructures grow in scale and complexity, intrusion detection and intrusion prevention systems (IDS/IPS) face a fundamental combinatorial challenge. The feature space describing network traffic is high-dimensional, and the combinations of features that signal an attack can be subtle and exponentially numerous. Traditional machine learning approaches to IDS rely on heuristic feature selection that achieves strong classification accuracy yet may miss interpretable, multi-feature interaction patterns specific to particular attack types. Even when these features perform well, their feature selection is typically embedded in black box architectures, which can limit cybersecurity analysts ability to understand why certain network connections are flagged, or how to determine how an attacker may have gotten through.

Subgroup Discovery can approach the network security surface from a different direction, by finding interpretable conjunctive rules that identify network connections that are unusual in respect to a target variable[1][2]. Subgroup discovery has been applied to a wide range of applications ranging from biomedical cohort prioritization [3] to network intrusion profiling [4], where Atzmueller et al. combined SD with complex network analysis to characterize attack patterns in an interpretable way in IDS data. This method can be used to improve current systems, however, finding optimal subgroups is a NP-hard combinatorial optimization problem where the number of subgroups to evaluate grows exponentially with the number of features. There are current classical heuristic algorithms that improve this scaling complexity by utilizing greedy pruning algorithms which can miss subgroups where the components may seem weak but together can lead to meaningful conjunctive rules.

In parallel, Noisy Intermediate-Scale Quantum (NISQ) hardware has continued to improve in qubit count, gate fidelity, and cloud accessibility, motivating researchers to explore applications where quantum algorithms may offer structural advantages over classical methods for combinatorially hard problems [5][6]. This project targets one such application, discovering subgroups of network connections whose specific combinations of feature values are strongly associated with attack traffic in the NSL-KDD benchmark. Identifying these subgroups is an NP-hard combinatorial optimization problem whose search space grows exponentially with the number of candidate features, making it a natural target for hybrid quantum-classical methods.

We formulate the subgroup discovery problem as a QUBO and solve it with QAOA on IBM Quantum hardware. A complete pipeline and unique QUBO formulation for increasing fidelity to the true WRAcc objective is shown, and the results are benchmarked against classical Beam Search and an exhaustive ground-truth enumeration when possible.

## II. BACKGROUND & RELATED WORK

## 2.1 Classical Subgroup Discovery

Subgroup discovery (SD) seeks compact, interpretable descriptions of data subsets that behave unusually with respect to a target variable [1][2]. For a binary attack label, the standard quality measure is Weighted Relative Accuracy (WRAcc):

$$WRAcc(S) = \left(\frac{|sg|}{N}\right) \cdot |p(sg) - p_0| \qquad (1)$$

where $S$ is a candidate subgroup, $|sg|$ is the number of records satisfying all selected feature conditions, $N$ is the total record count, $p(sg)$ is the positive (attack) rate within the subgroup, and $p^0$ is the overall positive rate over all $N$ records. WRAcc rewards both coverage (reaching many records) and contrast (deviating markedly from the baseline rate), preventing trivially small or trivially large subgroups [1][2].

Lavrač et al. [2] established the foundational WRAcc-based SD framework with the CN2-SD algorithm, which adapts rule learning for subgroup identification. Kavšek et al. [1] extended this line with APRIORI-SD, adapting association rule learning techniques to subgroup discovery and demonstrating that different search strategies uncover complementary subgroup structures. The dominant practical algorithm remains beam search: starting from individual features, the search iteratively extends subgroups by adding one feature at a time, retaining only the top-scoring beam_width candidates by WRAcc at each depth. This greedy pruning makes beam search efficient but can miss subgroups whose individual features appear weak yet are jointly discriminative. Exhaustive enumeration of all C(n,k) subsets guarantees the optimum but is feasible only for small n and k, serving here as the ground-truth baseline.

SD has been applied beyond traditional data mining contexts. Liu et al. [3] developed an exploratory subgroup cohort discovery method for biomedical research, demonstrating SD's value for identifying interpretable contrast patterns in high-dimensional clinical data. Most relevant to this work, Atzmueller et al. [4] applied SD combined with complex network analysis directly to network intrusion profiling, establishing that SD is a viable and valuable paradigm for characterizing attack patterns in IDS data. The present work extends this direction by bringing quantum combinatorial optimization to the SD search process itself.

## 2.2 Network Intrusion Detection and Prevention

The NSL-KDD dataset [7] is the standard benchmark for evaluating network IDS methods. It corrects critical shortcomings of the original KDD Cup 1999 dataset by removing redundant records that bias classifiers toward frequent patterns. Each record describes one network connection with 41 features and a string attack label. The standard evaluation protocol uses KDDTrain+ for training and KDDTest+ for held-out evaluation, with KDDTest-21 providing a harder subset that excludes easily classified records. Four attack families are defined: Denial of Service (DoS), Probe, Remote-to-Local (R2L), and User-to-Root (U2R).

In the classical IDS literature, machine learning approaches from decision trees to deep learning have been applied to NSL-KDD with strong classification accuracy. Dhanabal and Shantharajah [8] provided an early systematic study of classification algorithms on NSL-KDD, while Sharafaldin et al. [9] offered a deterministic comparison of classical and hybrid deep representation models across NSL-KDD and CICIDS2017. These approaches optimize prediction metrics (accuracy, F1, AUC) rather than interpretable subgroup quality, and their feature selection is typically embedded within black-box models. Feature interaction is particularly important for IDS: different attack types exploit different feature combinations that may appear benign in isolation, a property that classification-oriented approaches are not designed to expose.

## 2.3 Quantum Computing in the NISQ Era

QAOA [10] is a hybrid quantum-classical variational algorithm for combinatorial optimization. A p-layer QAOA circuit prepares the variational state defined as the following function:

$$|\psi(\gamma,\beta)\rangle = \prod_{l=1}^{p} e^{-i\beta_l B}\, e^{-i\gamma_l H_C} \,|+\rangle^{\otimes n} \qquad (2)$$

where p is the number of QAOA layers, $\boldsymbol{\gamma} = (\gamma_1, \ldots, \gamma_p)$ and $\boldsymbol{\beta} = (\beta_1, \ldots, \beta_p)$ are the variational angle parameters, H_C is the cost Hamiltonian (encoding the QUBO objective), $B = \Sigma_i X_i$ is the transverse-field mixer Hamiltonian, and $|+\rangle^{\otimes n}$ is the equal-superposition initial state on n qubits. A classical outer-loop optimizer (here COBYLA) minimizes the expectation value $\langle\psi(\gamma,\beta)|H_c|\psi(\gamma,\beta)\rangle$ and the resulting bitstring probability distribution is sampled to recover candidate feature subsets.

The QUBO-to-Ising mapping converts binary variables $x_i \in \{0,1\}$ to Ising spins $z_i \in \{-1,+1\}$ via $x_i = \frac{1-z_i}{2}$. Under this transformation, the QUBO objective $x^T Q x$ becomes the Ising cost Hamiltonian:

$$H_C = \sum_i h_i Z_i + \sum_{i<j} J_{ij} Z_i Z_j + c \qquad (3)$$

where $Z_i$ is the Pauli-Z operator on qubit $i$, $h_i$ are the local field coefficients derived from the diagonal of $Q$, $J_{ij}$ are the coupling coefficients derived from the off-diagonal of $Q$, and $c$ is a constant offset that does not affect the ground state. Both $h_i$ and $J_{ij}$ map directly to single-qubit Z and two-qubit ZZ gates implementable on superconducting hardware.

Blunt et al. [5] surveyed quantum algorithms applicable to complex optimization and identified variational methods such as QAOA as the most promising near-term approach. Omanakuttan et al. [6] derived fault-tolerance thresholds for quantum advantage with QAOA on optimization tasks, providing analytical crossover-point estimates that inform the scalability analysis in this work. In a related data mining context, Yu [11] experimentally implemented quantum association rule mining (qARM) on IBM quantum hardware, demonstrating that quantum circuits can successfully recover frequent itemsets from transaction databases, the closest existing precedent for applying quantum computation to a combinatorial data mining task analogous to subgroup discovery. Li and Wang [12] developed a collaborative neurodynamic algorithm as an alternative classical solver for QUBO problems, providing context for the broader landscape of QUBO solution methods against which quantum approaches must be evaluated.

## 2.4 Quantum Machine Learning for Intrusion Detection

A growing body of work has applied quantum and hybrid quantum-classical models directly to network intrusion detection as classification tasks. Gouveia and Correia [13] provided early evidence by evaluating quantum-enhanced classifiers on network traffic data at IEEE NCA 2020, establishing a foundation for subsequent QML-IDS research. Gong et al. [14] developed an attack detection scheme based on variational quantum neural networks, demonstrating improved accuracy over classical baselines for certain attack types. Kukliansky et al. [15] evaluated quantum neural networks for network anomaly detection on actual noisy quantum hardware, providing an important NISQ-era benchmark for the practical fidelity of quantum IDS models. Wang et al. [16] proposed a hybrid quantum-enhanced convolutional neural network for network attack traffic detection, showing that quantum layers can improve feature extraction from traffic data.

Abreu et al. [17], [18] built QML-IDS and QuantumNetSec systems using quantum support vector machines on the NSL-KDD dataset, achieving competitive classification performance and further establishing NSL-KDD as the standard testbed for quantum IDS research. Kadi et al. [19] conducted a systematic comparative study of quantum-classical encoding methods for network intrusion detection, providing insights into how data representation choices affect quantum classifier performance. Anguera et al. [20] extended QML-IDS approaches to software-defined networking environments. Bellante et al. [21] critically evaluated the potential of quantum machine learning for cybersecurity through a case study on PCA-based intrusion detection, explicitly assessing whether current quantum methods provide an advantage over classical alternatives. The motivation of this question is related to the decision for dual approximation-ratio framework adopted in this work.

It is important to highlight that all of the above QML-IDS work targets *classification accuracy*, given a network connection, predict whether it is an attack. None formulates the problem as combinatorial optimization over feature subsets to discover *interpretable subgroups* that characterize attack patterns. This paper has a similar motivation but attempts to solve it a different way, by casting subgroup discovery as a QUBO and solving it with QAOA, applying quantum combinatorial optimization directly to the cybersecurity feature-interaction problem and evaluating results against both a classical heuristic baseline and an exhaustive ground-truth optimum.

*2.5 Combinatorial Optimization for Feature Selection*

Feature selection for IDS is itself a combinatorial optimization problem: given $n$ candidate features, the goal is to identify a subset of size $k$ that optimizes some quality criterion, requiring a search over $C(n,k)$ possible subsets. Classical approaches span a spectrum from filter methods (ranking features independently by information gain (IG) or mutual information) to wrapper and embedded methods that evaluate subsets jointly but rely on heuristic search to avoid exhaustive enumeration.

In the IDS domain, metaheuristic optimization has become a popular strategy for navigating this combinatorial space. Ghanbarzadeh et al. [22] developed a quantum-inspired multi-objective horse herd optimization algorithm (MQBHOA) for feature selection on NSL-KDD and CSE-CIC-IDS2018, integrating quantum-inspired qubit and quantum gate operators into the search process to balance exploration and exploitation. Logeswari et al. [23] proposed a hybrid feature selection pipeline combining filter and wrapper stages for IoT intrusion detection, demonstrating that multi-stage approaches can reduce feature dimensionality while maintaining classification accuracy. This work is meaningful to quantum implementations where current hardware limits the number of features. Turaka and Panigrahy [24] combined quantum diffusion modeling with chaos theory and deep learning for feature optimization in IoT IDS, achieving up to 75% feature reduction on NSL-KDD and related benchmarks. Barati [25] integrated a quantum genetic algorithm with self-supervised learning for lightweight feature selection in resource-constrained IoT and wireless sensor network environments. Shen et al. [26] presented a quantum-inspired computing approach to IDS that uses quantum-inspired representations to improve the efficiency of the detection pipeline.

Iovane [27] provided a broader survey of quantum-inspired algorithms for cybersecurity optimization, documenting how quantum evolutionary algorithms and quantum kernel methods can enhance classical optimization pipelines for tasks including intrusion detection. Li and Wang [12] developed a collaborative neurodynamic algorithm as a classical solver for QUBO problems, offering an alternative optimization paradigm for the same class of binary optimization formulations used in this work.

The explainability criteria separates all of the above from the present work. The metaheuristic and quantum-inspired approaches treat feature selection as a *preprocessing step* for a downstream classifier: the selected features are fed into KNN, SVM, or deep learning models, and success is measured by classification accuracy. The feature subset itself has no independent interpretive value. In contrast, this paper formulates feature *subset* discovery as the primary objective, where the selected features directly define an interpretable subgroup rule evaluated by WRAcc. Furthermore, while the quantum-inspired methods above use classical hardware with quantum-motivated algorithmic components (Qubits, quantum gates as metaphors for search operators), this work executes QAOA on actual quantum hardware, encoding the feature selection objective as a QUBO Hamiltonian solved through quantum superposition and entanglement on IBM Quantum superconducting processors.

Three gaps in this literature motivate our approach. First, no prior quantum work on NSL-KDD applies optimization to the discovery of interpretable rules, the surveyed methods all target classification accuracy or use feature selection as a preprocessing step. Second, the quantum-inspired metaheuristics (MQBHOA, quantum genetic algorithms, horse-herd) run on classical hardware and have never been compared against an exhaustive optimum on real quantum devices. Third, the QAOA-on-IDS literature offers scaling projections but no measured fidelity boundary for dense QUBO instances on production hardware. We address each below.

III. METHODOLOGY

This section presents the pipeline that addresses those three gaps. We encode WRAcc as a least-squares-fit QUBO so that subgroup quality becomes a quantity QAOA can directly optimize, with an Ising form whose ZZ couplings are non-trivial enough to actually entangle qubits during the cost layer. We benchmark against both beam search (the standard heuristic) and exhaustive enumeration where it remains tractable, scoring QAOA outputs on Hamiltonian quality ($r_E$) and on the application metric we actually care about, recovered WRAcc ($r_W$). And we run the whole pipeline on ibm_pittsburgh from 10 to 30 qubits to find where current hardware stops being useful. The output is an interpretable rule set for analysts and a measured, not estimated, picture of what dense-QUBO QAOA can do at this hardware generation.

The pipeline consists of five sequential phases: (1) data preprocessing, (2) classical baselines, (3) QUBO construction, (4) QAOA execution, and (5) evaluation. Algorithm 1 provides a pseudocode summary; each phase is detailed below.

### 3.1 Phase 1: Data Acquisition and Preprocessing

The NSL-KDD dataset is loaded from its standard headerless text format, where each record consists of 41 features, a string attack-type label, and a difficulty score. Attack labels are mapped to a binary is_attack target (0 for normal, 1 for any attack) with additional attack category labels (DoS, Probe, R2L, U2R).

<u>Categorical encoding.</u> Three categorical features, protocol_type, service, and flag, undergo one-hot encoding, with cardinality-aware handling introduced to manage qubit budgets. Low-cardinality categoricals (≤5 unique values, e.g., protocol_type with 3 values) receive full one-hot encoding. High-cardinality categoricals (>5 unique values, notably service

```
ALGORITHM 1: Quantum-Enhanced Subgroup Discovery Pipeline

Input:  NSL-KDD dataset D, target cardinality K,
        attack filter a ∈ {all, DoS, Probe, R2L, U2R},
        binary side s ∈ {0, 1, both},
        top-K_feat features, QAOA depths P = {p₁, ..., p_m}
Output: Ranked subgroups, approximation ratios, QAOA-unique
discoveries
//Phase 1: Data Preprocessing
 1. Load D_train, D_test from KDDTrain+, KDDTest+
 2. One-hot encode categorical features (protocol_type, service, flag)
 3. IF a ≠ all THEN
 4.     Filter to retain only normal and attack category a
 5. X ← StandardScaler(D_train)
 6. IF s = 1 THEN
 7.     B ← 𝟙[X > 0]
 8. ELSE IF s = 0 THEN
 9.     B ← 𝟙[X ≤ 0]
10. y ← 𝟙[attack_type ≠ normal]
11. Rank features by information gain; select top K_feat → n qubits
//Phase 2: Baselines
12. S_exh ← ExhaustiveEnum(B, y, K)
13. S_beam ← BeamSearch(B, y, width, max_depth)
14. WRAcc* ← max WRAcc(S) for S ∈ S_exh
//Phase 3: QUBO Construction
15. Q_obj ← LS-RegressionQUBO(B, y, K)
16. λ ← CalibratePenalty(Q_obj, K)
17. Q ← Q_obj + λ · CardinalityPenalty(K)
18. Q ← Q / max|Q|
19. H_C ← QUBOtoIsing(Q)
//Phase 4: QAOA Execution
20. FOR each depth p ∈ P DO
21.     R_p ← RunQAOA(H_C, p, n, backend)
22. END FOR
//Phase 5: Evaluation
23. FOR each result R_p DO
24.     Post-select bitstrings with |S| = K
25.     r_E ← E_QAOA / E_ground
26.     r_W ← WRAcc_best^(k=K) / WRAcc
27.     Evaluate WRAcc on D_test for generalization
28. END FOR
29. U ← {S ∈ R : S ∉ S_beam}
30. RETURN S_exh, S_beam, {R_p}, U, r_E, r_W
```

with 70 values) undergo information-gain filtering: per-category IG is computed against the binary attack label, categories are ranked by IG descending, and only those categories are retained until their cumulative IG accounts for ≥90% of the total column IG. Remaining categories collapse into an implicit all-zeros reference class, preserving full interpretability (rules like service_http = 1 remain intact) while strictly limiting the number of dummy columns potentially used for qubit allocation.

<u>Attack filtering.</u> The pipeline supports per-category attack analysis through an attack filter configuration. When set to a specific category (e.g., R2L), the data is restricted to records labeled normal plus records from that attack category, and the binary target is recomputed accordingly. When set to 'all', the original binary target (any attack vs. normal) is used. This filtering is applied to both training and test data.

<u>Standardization and binarization.</u> All features are standardized to zero mean and unit variance using a StandardScaler fit exclusively on training data. The binarization step converts each scaled feature to {0, 1} via thresholding. Two threshold strategies are supported: (1) mean thresholding, where the split point is at zero in scaled space (i.e., the training mean), and (2) entropy-optimal thresholding, where for each feature, up to 200 evenly spaced candidate split points between the feature's minimum and maximum scaled values are evaluated, and the threshold maximizing information gain with respect to the attack label is selected.

<u>Directional alignment.</u> When directional alignment is enabled, the pipeline determines per-feature polarity rather than applying a uniform binarization direction. For each feature, both threshold directions (value > threshold and value ≤ threshold) are evaluated. The direction whose subgroup has a higher attack rate than the overall baseline is assigned binary = 1, ensuring that a 1 consistently means "more anomalous / more attack-like" across all features. Since WRAcc is symmetric by construction (both directions yield identical WRAcc due to the absolute value on contrast), this alignment does not change the optimization landscape but ensures semantic consistency in the resulting subgroup rules. The polarity map and thresholds are stored from training and reapplied to test data to guarantee alignment.

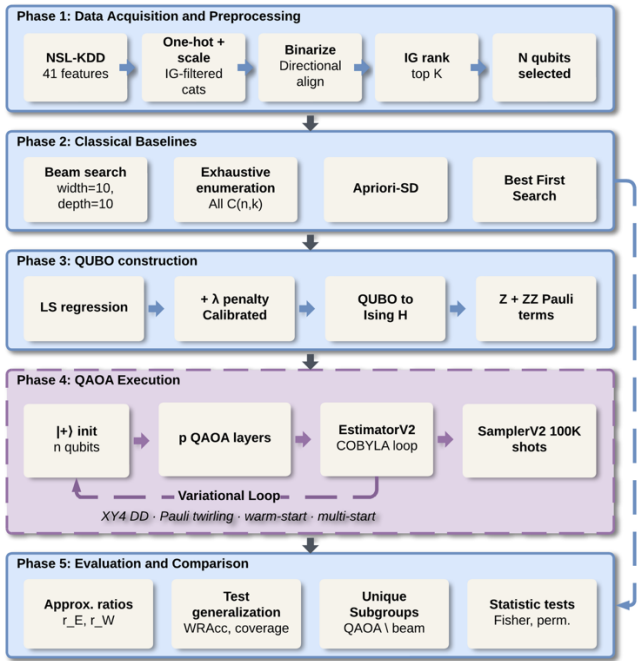

*Figure 1: End-to-end QAOA subgroup-discovery pipeline. Phases 1-2 binarize NSL-KDD features and run classical baselines. Phase 3 fits the LS-regression QUBO. Phase 4 executes QAOA on ibm_pittsburgh with COBYLA optimization and error mitigation. Phase 5 computes the dual approximation ratios and identifies QAOA-unique subgroups.*

Feature importance ranking. After binarization, features are ranked by their information gain with respect to the binary attack label. For a binarized feature $f$ taking values in $\{0, 1\}$ and target variable $Y$ taking values in $\{0, 1\}$, information gain is defined as the following function:

$$IG(f) = H(Y) - H(Y|f) \quad (4)$$

where the conditional entropy is defined as:

$$H(Y|f) = \sum_{v \in \{0,1\}} P(f = v) \cdot H(Y|f = v) \quad (5)$$

where $H(\cdot)$ is the binary Shannon entropy in bits, $P(f = v)$ is the probability that feature f takes value $v$ in the training set, and $H(Y | f = v)$ is the entropy of the attack label $Y$ restricted to the subset of records where $f = v$. The top-$K$ features by IG(f) are selected for the QUBO formulation ($K = 10$ in the primary R2L configuration, $K = 20$ in the general all-attack configuration). This step ensures each qubit represents a discriminative feature while keeping qubit requirements within NISQ hardware limits.

Evaluation modes. The pipeline supports four evaluation modes: (1) full: train on KDDTrain+.txt, evaluate on KDDTest+.txt (default); (2) 20pct: train on the 20% subset; (3) train_only: held-out evaluation on training data for rapid iteration; and (4) hard: evaluate on KDDTest-21.txt, a harder subset that excludes easily classified records. All reported results use full mode. The training set contains 125,973 records (46.5% attack) and the test set 22,544 records (56.9% attack) before attack filtering.

*3.2 Phase 2: Classical Baselines*

Two classical baselines establish the reference against which QAOA results are evaluated: an exhaustive enumeration providing the ground-truth optimum, and a beam search providing the primary heuristic comparator.

Exhaustive enumeration. For each cardinality k from 1 up to min(8, n), the pipeline enumerates all C(n, k) feature subsets and computes their WRAcc. For a given feature subset defined by binary selection vector $x$, a network connection is a subgroup member if and only if all selected binarized features equal 1, that is, the subgroup is the conjunction rule defined by $x$. WRAcc for each subset is computed using Eq. (1). A safety threshold caps exhaustive enumeration at 50,000 combinations per cardinality to prevent intractable computation at larger $n$; cardinalities exceeding this threshold are skipped with a warning. The best WRAcc is then used as the denominator of the WRAcc approximation ratio.

Beam search. A beam search with configurable width (BEAM_WIDTH, default 10) and maximum depth (BEAM_MAX_DEPTH, default 10) serves as the primary classical heuristic baseline. Starting from all n individual features as depth-1 candidates, the search iteratively extends subgroups by adding one feature at a time. At each depth, only the top BEAM_WIDTH subgroups by WRAcc are retained as seeds for the next level. Candidate propagation accumulates extension results after the full inner loop completes, eliminating duplicate candidates. All discovered subgroups across all depths are collected, deduplicated, and sorted by WRAcc.

Additional SD baselines. When available, the pipeline also runs standard subgroup discovery algorithms from the pysubgroup library as supplementary baselines: Best-First Search (BSD) [28], Apriori-SD [1], SimpleDFS, and pysubgroup's own beam search implementation [29], all using WRAcc as the quality function. These provide additional reference points but are not used as primary comparators in the approximation ratio calculations due to similar performance.

All discovered subgroups, from exhaustive enumeration, beam search, and supplementary baselines, are evaluated for WRAcc, coverage, and contrast on both training and test data. Beam search recall is reported as the number of the exhaustive top-20 subgroups at target cardinality that also appear in the beam search results, quantifying how much of the true optimum the heuristic recovers.

*3.3 Phase 3: QUBO Construction*

Least-squares regression QUBO. The subgroup discovery objective is encoded as a Quadratic Unconstrained Binary Optimization (QUBO) problem via least-squares regression, producing a quadratic function $x^T Q_x$ that approximates the negated WRAcc landscape over feature subsets. The fitting objective is:

$$Q^* = argmin_Q \sum_{x \in S} \left(x^T Q x - (-WRAcc(x))\right)^2 \quad (6)$$

where $Q*$ is the fitted QUBO matrix, $x \in \{0,1\}^n$ is a binary feature-selection vector, S is the set of training bitstrings used for the regression (described below), and the negation of $WRAcc(x)$ reflects that QUBO minimization corresponds to WRAcc maximization. The QUBO matrix has $n$ diagonal entries (one per feature) and $\frac{n(n-1)}{2}$ off-diagonal entries (one per feature pair), giving $V = n + \frac{n(n-1)}{2}$ total variables in the regression. The design matrix $A$ encodes each sampled bitstring as a row vector of length $V$: the first $n$ entries are the binary feature selections $x_i$, and the remaining entries are the pairwise products $2 \cdot x_i \cdot x_j$ for each $i < j$. The target vector contains the negated WRAcc values.

Two construction modes are supported. For small feature counts (n ≤ 15), exact enumeration computes WRAcc for all 2^n bitstrings and restricts the regression to k = TARGET_FEATURES bitstrings. For larger feature counts (n > 15), surrogate sampling generates M = SURROGATE_OVERSAMPLE × V randomly sampled bitstrings (with balanced Hamming weight sampling to ensure coverage across cardinalities) and fits Q to this sample. The surrogate approach reduces QUBO construction from O(2^n) to O(n²) while preserving the essential structure of the WRAcc landscape. Fit quality is reported via R² and Spearman rank correlation ρ. Because the LS-regression naturally produces non-uniform off-diagonal coefficients, this formulation reliably generates two-qubit entangling gates (ZZ terms) after transpilation, a critical requirement for QAOA circuits to produce non-trivial correlations between qubits.

Pre-normalization. Before adding the cardinality penalty, the objective QUBO Q_obj is normalized by max|Q_obj|. This preserves the signal-to-penalty ratio so that QAOA can distinguish between different k-subsets rather than seeing a landscape dominated by the penalty term.

Cardinality penalty calibration. To encourage QAOA to select exactly K features, an additive cardinality penalty is included, defined as:

$$P_{card}(x) = \lambda \left( \sum_i x_i - K \right)^2 \quad (7)$$

where $\lambda > 0$ is the penalty weight calibrated as described below, and the squared term vanishes only when the cardinality constraint $\sum_i x_i = K$ is satisfied exactly. The penalty weight $\lambda$ is calibrated analytically: for each candidate wrong-cardinality $k \neq K$, the pipeline computes the best objective energy at that k and determines the minimum penalty needed to make the target-cardinality solution favorable. The final $\lambda$ is set to 1.5× the maximum needed penalty across all candidate wrong-k values, with a minimum floor of 0.1. Unlike fixed-penalty heuristics, this per-instance calibration ensures the cardinality constraint is tight without over-penalizing the energy landscape, which would suppress the WRAcc signal QAOA must navigate. Expanding the squared term and using $x_i^2 = x_i$ for binary variables, the penalty contributes the following terms to the QUBO:

$$P_{card}(x) = \lambda \left[ (1 - 2K) \sum_i x_i + 2 \sum_{i<j} x_i x_j + K^2 \right] \quad (8)$$

where the $(1 - 2K)$ coefficient adds to each diagonal entry of $Q$, the factor of $2\lambda$ is added to each off-diagonal pair, and $\lambda K^2$ is a constant offset that does not affect the optimizer. When FREE_CARDINALITY mode is enabled, no cardinality penalty is added, allowing QAOA to find the best WRAcc at any k. In this mode, the WRAcc approximation ratio compares against the global exhaustive optimum across all cardinalities.

Final normalization and Ising conversion. The penalized QUBO matrix is normalized by dividing by max|Q| to bring all coefficients into a numerically stable range. The QUBO-to-Ising mapping then converts each binary variable $x_i \in \{0, 1\}$ to an Ising spin $z_i \in \{-1, +1\}$ via the substitution:

$$x_i = \frac{1 - z_i}{2} \quad (9)$$

Applying this substitution, the objective $x^T Q x$ takes the Ising form already given in Eq. (3), with $h_i$ and $J_{ij}$ derived directly from the entries of $Q$. QUBO/Ising consistency is verified by evaluating random bitstrings under both representations and confirming agreement to within numerical precision.

*3.4 Phase 4: QAOA Execution*

The Quantum Approximate Optimization Algorithm is used to find low-energy states of the cost Hamiltonian $H_C$ (Eq. (3)), which correspond to high-WRAcc feature subsets. A p-layer QAOA circuit prepares the variational state given in Eq. (2), alternating applications of $H_C$ and the transverse-field mixer Hamiltonian $B$ over $p$ layers starting from an equal superposition on $n$ qubits.

Circuit construction and transpilation. The QAOA circuit is constructed using Qiskit's QAOAAnsatz with measurement operations appended. For simulator runs, transpilation uses optimization level 1 with a linear coupling map; for hardware runs, optimization level 3 is used with the target backend's native topology. Gate counts are reported after transpilation, including two-qubit gate count (CX/ECR/CZ), total gate count, and circuit depth. A warning is raised if zero two-qubit gates are produced despite non-zero ZZ terms in the Hamiltonian, indicating a potential transpilation issue.

Variational optimization. The COBYLA gradient-free optimizer minimizes the QAOA cost function:

$$F(\gamma, \beta) = \langle \psi(\gamma, \beta) | H_C | \psi(\gamma, \beta) \rangle \quad (10)$$

where $|\psi(\gamma, \beta)\rangle$ is the QAOA variational state from Eq. (2), and the expectation is evaluated via IBM Quantum's EstimatorV2 primitive at each optimizer iteration. Two convergence enhancements are applied:

- Warm-start. Optimal parameters from a depth-p run initialize the depth-(p+1) optimization, propagating good solutions across circuit depths. If the parameter count increases (2p → 2(p+1)), additional parameters are initialized with small random values near zero.

- Multi-start. N_OPTIM_RESTARTS random initial parameter vectors are tried in each depth run and the best result is retained, reducing sensitivity to local optima. The first restart uses the default initialization ($\frac{\pi}{4}$ for all parameters, or warm-start parameters if available); subsequent restarts use uniformly random parameters in $[0, \pi]$.

For hardware runs, each COBYLA iteration creates a fresh EstimatorV2 instance and submits an independent job to the IBM Quantum cloud (the separate-jobs architecture), avoiding the continuous Session billing that accumulates during variational loops. This architecture trades increased job submission latency for reduced QPU-time cost, an important practical consideration for iterative variational algorithms on cloud-accessible hardware.

Error mitigation. Two error mitigation strategies are applied on hardware:

- Dynamical decoupling with the XY4 pulse sequence to suppress idle-qubit decoherence during the idle periods between gate operations.

- Pauli gate twirling (num_randomizations = 'auto') to convert coherent errors into stochastic noise that is more amenable to post-processing and averaging.

Sampling. After convergence, the optimized circuit parameters are fixed and a final SamplerV2 job samples the circuit with SAMPLER_SHOTS (default 100,000) shots to produce a probability distribution over bitstrings. The resulting measurement counts are converted to a probability distribution over integer-valued bitstring indices.

*3.5 Phase 5: Evaluation and Comparison*

The top-N most probable bitstrings (default N = 50) from the QAOA sampling distribution are decoded into feature subsets and evaluated comprehensively.

Subgroup metrics. Each decoded bitstring defines a feature subset, and the corresponding subgroup is the set of records satisfying all selected binarized features simultaneously (conjunction rule). For each subgroup, WRAcc, coverage (fraction of records captured), contrast (absolute difference between subgroup positive rate and overall positive rate), and positive rate are computed on both training and test data.

Dual approximation ratios. Two approximation ratios provide complementary evaluations of QAOA performance:

The *energy approximation ratio* $r_E$ compares QAOA's best sampled Ising energy against the true ground-state energy and is defined as:

$$r_E = \frac{\langle H_C \rangle_{QAOA}}{E_{ground}} \quad (11)$$

where $\langle H_C \rangle_{QAOA}$ is the lowest Ising energy among the QAOA-sampled bitstrings (evaluated under the cost Hamiltonian $H_C$), and E_ground is the true ground-state energy of $H_C$ obtained by exhaustive enumeration over the $2^n$ bitstrings. By construction, $r_E \leq 1$, with $r_E = 1$ indicating recovery of the exact ground state.

We define a second ratio, $r_W$, to measure solution quality directly in the subgroup discovery objective. It compares the best post-selected QAOA WRAcc at the target cardinality against the exhaustive optimum:

$$r_W = \frac{WRAcc_{best}^{(k=K)}}{WRAcc^*} \quad (12)$$

where $WRAcc_{best}^{k=K}$ is the highest WRAcc among QAOA-sampled bitstrings whose Hamming weight equals the target cardinality K, and $WRAcc*$ is the exhaustive optimum WRAcc at cardinality K over all $C(n,K)$ feature subsets. Values $r_W \geq 1$ are possible when the unconstrained best bitstring has $k \neq K$ and QAOA samples high-quality bitstrings across multiple cardinalities. When FREE_CARDINALITY mode is enabled, $WRAcc*$ is replaced by the global exhaustive optimum across all k.

Statistical validation. For subgroups with sufficient membership, two statistical tests assess significance:
  - Fisher's exact test on the 2×2 contingency table (in-subgroup vs. outside, attack vs. normal) provides an odds ratio and p-value for the association between subgroup membership and attack status.
  - A permutation test (1,000 permutations) shuffles attack labels, recomputes WRAcc for each permutation, and reports the fraction of permuted WRAccs exceeding the observed value as an empirical p-value. A z-score quantifies how many standard deviations the observed WRAcc exceeds the null distribution mean.

QAOA-unique subgroup identification. The pipeline identifies QAOA-unique subgroups as those feature combinations present in the QAOA sampling distribution but absent from beam search results. These represent regions of the combinatorial search space that greedy heuristics prune, providing evidence of the quantum pipeline's ability to surface feature interactions that classical methods miss.

Test-set generalization. All subgroups, from beam search, exhaustive enumeration, and QAOA, are evaluated on the held-out KDDTest+ set using the same binarization thresholds and polarity map learned from training data. Test-set WRAcc, coverage, contrast, and positive rate provide evidence of generalization rather than training-set overfitting.

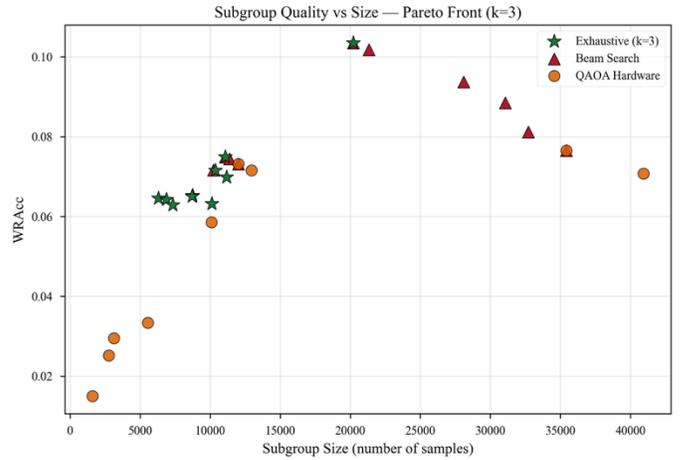

*Figure 2: Subgroup quality vs. subgroup size at k = 3 for the 20-qubit Probe configuration on ibm_pittsburgh. Green stars mark the exhaustive optimum (best WRAcc at any k), red triangles the beam-search results, and orange diamonds the QAOA-sampled subgroups. QAOA recovers subgroups in the same Pareto region as the classical baselines, confirming hardware competitiveness at this scale ($r_W = 0.855$).*

IV. RESULTS

*4.1 Classical Beam Search Results*

Table I presents beam search performance on the full NSL-KDD training set, with corresponding QAOA hardware results. For context, in the aggregate all-attack setting (not shown in Table I), the best classical subgroup discovered by beam search is the two-feature combination {flag, same_srv_rate} (train $WRAcc = 0.1798$), capturing connections with non-standard TCP flags and low service homogeneity as strongly associated with attack traffic. Results for this configuration are consistent across train and test splits, confirming the validity of the binarization and WRAcc framework.

Beam search recall of the exhaustive top-20 subgroups at target cardinality varies by attack category. For the R2L-filtered configuration at 10 qubits with $k = 6$, beam search recalls 16 of the 20 exhaustive-optimal subgroups, leaving 4 subgroups that greedy pruning misses. This recall gap widens at larger $n$: at 20 qubits, the search space grows from C(10,6) = 210 to C(20,6) = 38,760 subsets, and the beam search miss rate increases correspondingly.

*4.2 QUBO Fit Quality*

The least-squares regression QUBO achieves $R^2 = 0.989$ and Spearman $\rho = 0.899$ between the quadratic objective and the true WRAcc landscape for the R2L-filtered 10-qubit configuration. The off-diagonal coefficients exhibit sufficient diversity (non-trivial standard deviation relative to their mean) to ensure that transpilation produces two-qubit entangling gates (ZZ terms) in the QAOA circuit. QUBO/Ising consistency is verified to within numerical precision across random bitstring samples. The QUBO ground state matches the exhaustive optimum at target cardinality $k = 6$, confirming that the quadratic approximation matches the WRAcc ranking at the target cardinality.

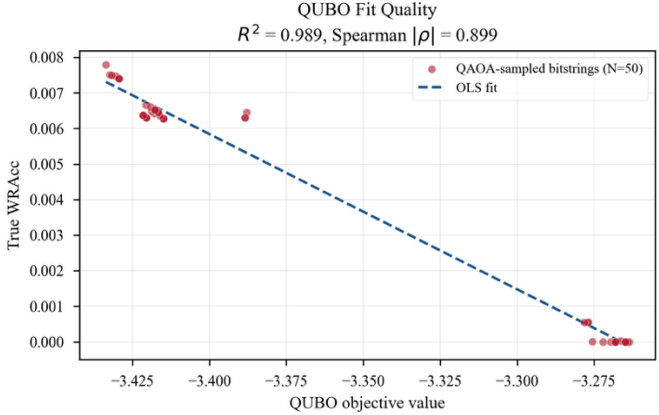

*Figure 3: QUBO fit quality for the R2L-filtered 10-qubit configuration. Each point is one QAOA-sampled bitstring plotted against its quadratic QUBO objective value; the OLS fit ($R^2 = 0.989$, Spearman $|\rho| = 0.899$) confirms the QUBO accurately ranks the WRAcc landscape.*

*4.3 QAOA Subgroup Discovery Results*

On the noiseless AerSimulator at 10 qubits (R2L filter, $k = 6$), QAOA at depth $p = 1$ achieves a WRAcc approximation ratio $r_W = 1.0$, recovering the exhaustive-optimal subgroup {dst_host_srv_count, srv_count, service_ftp_data, service_http, dst_host_same_src_port_rate, count} with WRAcc = 0.00779 (train) and test $WRAcc = 0.0218$. The best QAOA subgroup achieves a training positive rate of 0.820 and a test positive rate of 0.957, meaning that when this six-feature conjunction rule fires, 95.7% of the matched test connections are confirmed R2L attacks. Increasing circuit depth to $p = 2$ maintains $r_W = 1.0$ while shifting more probability mass toward the target cardinality: probability at $k = 6$ increases from 11.8% ($p = 1$) to 13.7% ($p = 2$), and probability within $\pm 1$ of $k$ increases from 12.5% to 16.1%. This simulator baseline uses $k = 6$ on the 10-qubit R2L feature set; the hardware experiments reported in Table I use $k = 5$, matching the exhaustive optimum cardinality for the corresponding 15-qubit R2L feature selection.

TABLE I
Quantum Hardware Runs

| Attack | Qubits | Size | Classical | Quantum | Ratio | Depth |
|---|---|---|---|---|---|---|
| Probe | 10 | k=3 | 0.103515 | 0.101770 | 0.983 | p=1 |
| R2L | 15 (run 1) | k=5 | 0.012139 | 0.011788 | 0.971 | p=1 |
| R2L | 15 (run 2) | k=5 | 0.012139 | 0.010610 | 0.874 | p=1 |
| Probe | 20 | k=3 | 0.103515 | 0.088487 | 0.855 | p=1 |
| R2L | 25 | k=5 | 0.012865 | 0.008029 | 0.624 | p=1 |
| Probe | 30 | k=3 | 0.110480 | 0.004285 | 0.039 | p=1 |

Table I summarizes QAOA performance on ibm_pittsburgh across qubit counts ranging from 10 to 30 and attack-category filters (Probe, R2L). At 10 qubits, QAOA at p = 1 recovers the exhaustive optimum to within 1.7% ($r_W = 0.983$ on Probe). Performance remains competitive through 15 qubits ($r_W = 0.971$ on R2L, first run) and 20 qubits ($r_W = 0.855$ on Probe), then degrades to $r_W = 0.624$ at 25 qubits (R2L) and drops to $r_W = 0.039$ at 30 qubits as circuit noise dominates. Run-to-run variance is observed at 15 qubits, with a second R2L run yielding $r_W = 0.874$, reflecting sensitivity to hardware calibration drift and COBYLA initialization across jobs. These results place the practical NISQ boundary for dense QUBO instances at $p = 1$ near 20-25 qubits on current superconducting hardware.

*4.4 QAOA-Unique Subgroups*

QAOA-unique subgroups are feature combinations present in the sampled bitstring distribution but absent from beam search top candidates; they represent the most substantive contribution of the quantum pipeline. Among the top 50 QAOA-sampled subgroups in the R2L configuration, several combinations involving dst_host_srv_diff_host_rate and dst_host_count with service indicators and connection-count features appear consistently across both $p = 1$ and $p = 2$ runs but are absent from beam search results. Table II presents the combined top-15 R2L subgroup rules by training WRAcc, filtered to positive rate > 0.7, with each rule annotated by source. The top four ranks are recovered by both methods, confirming that the QUBO encoding correctly prioritizes the highest-WRAcc subgroups. Ranks 6, 7, and 10 are QAOA-unique, combinations involving dst_host_srv_diff_host_rate paired with service and host-count features that beam search does not find. Among the lower-ranked rules (11-15), beam search recovers simpler host-based combinations independently, while QAOA returns the same rule 13 and a unique rule at 15.

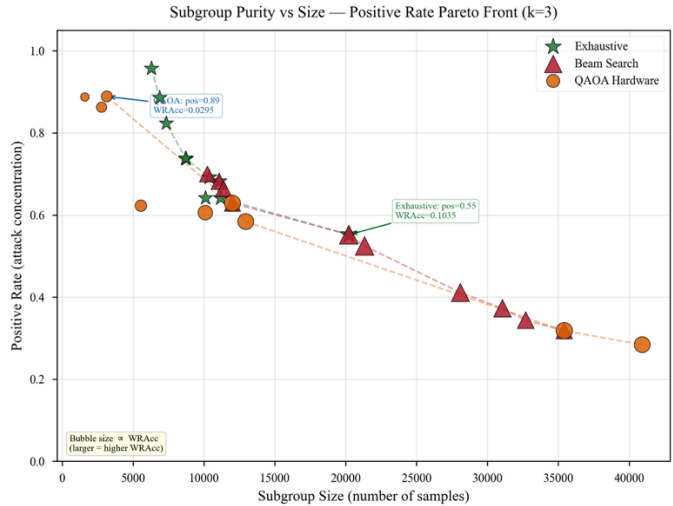

*Figure 4: Bubble Pareto front (k = 3) for the 20-qubit Probe configuration on ibm_pittsburgh. Bubble size encodes WRAcc. QAOA hardware subgroups are competitive with classical methods, with several returning subgroups exhibiting higher positive rates than beam search.*

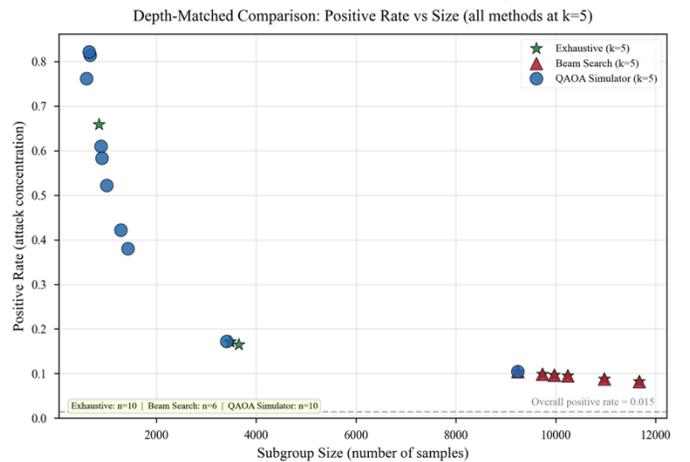

*Figure 5: Depth-matched comparison at $k = 5$ for the R2L-filtered configuration. The QAOA simulator returns smaller subgroups with higher attack-positive rate than beam search at the same cardinality, evidence that the QUBO objective rewards contrast-rich rules even when coverage is reduced.*

The attack-category breakdown reveals that these unique subgroups capture R2L signatures with high specificity. The six-feature subgroup {service_ftp_data, srv_count, count, service_http, dst_host_srv_diff_host_rate, dst_host_count} achieves a 91.6% R2L attack rate within the training subgroup (489 connections) and 99.6% precision on the test set (281 matched connections). Multiple QAOA-unique subgroups achieve 100% test precision (all matched connections are confirmed attacks), in this case with smaller subgroup sizes (12-22 connections). These patterns combine service diversity (dst_host_srv_diff_host_rate), connection-volume features (count, srv_count), and host-behavior metrics (dst_host_count) in ways that beam search does not reach because each component feature has only moderate individual WRAcc.

This result shows why the two search strategies differ. Beam search extends subgroups one feature at a time, retaining only the top-scoring candidates at each depth. If a two-feature intermediate has low WRAcc, it is pruned before the critical third or fourth feature can be added. QAOA's superposition-based search evaluates the full combinatorial space simultaneously, allowing it to approximate multi-feature interaction patterns where the components may appear weak in isolation.

TABLE II

Combined Top-8 R2L Subgroup Rules by Training WRAcc Filtered by PosRate >0.7

| Rank | WRAcc | Source | Features |
|---|---|---|---|
| 1 | 0.0079 | Both | count, dst_host_same_src_port_rate, service_ftp_data |
| 2 | 0.0078 | Both | dst_host_same_src_port_rate, service_ftp_data, srv_count |
| 3 | 0.0075 | Both | dst_host_count, dst_host_same_src_port_rate, service_ftp_data |
| 4 | 0.0067 | Both | count, dst_host_srv_diff_host_rate, service_ftp_data |
| 5 | 0.0067 | Classical | dst_host_srv_diff_host_rate, service_ftp_data |
| 6 | 0.0067 | *Quantum* | dst_host_srv_diff_host_rate, service_ftp_data, service_http |
| 7 | 0.0067 | *Quantum* | dst_host_srv_count, dst_host_srv_diff_host_rate, service_ftp_data |
| 8 | 0.0067 | Both | dst_host_srv_diff_host_rate, service_ftp_data, srv_count |

*4.5 Scaling Analysis*

We characterize how QAOA's recovery of the exhaustive optimum scales with qubit count n on ibm_pittsburgh, where C(n,k) grows combinatorially while QAOA's variational state explores the full $2^n$ space in superposition.

Scaling experiments varying n from 10 to 30 qubits reveal the impact of noise on results (Fig 6). As n increases, the classical exhaustive search space grows combinatorially, from 120 subsets at C(10,3) to 1,140 at C(20,3), while Beam Search's beam-recall of the exhaustive top-20 declines from 16/20 at $n = 10$ to lower values at larger $n$. In contrast to theoretical predictions that $p = 1$ QAOA performance should remain bounded independent of n, the measured hardware WRAcc ratio declines with qubit count: 0.983 at 10 qubits, 0.971 at 15, 0.855 at 20, 0.624 at 25, and 0.039 at 30 (Fig. 6, left). This degradation is driven primarily by noise accumulation in the transpiled circuit, tracking closely with the collapse in circuit-level fidelity floor (Fig. 6, right), rather than by the variational optimization itself, since the noiseless simulator retains $r_W = 1.0$ across this range. The crossover point where QAOA-unique subgroups could become reproducibly competitive with Beam Search is therefore gated by hardware fidelity rather than by the variational algorithm's theoretical reach.

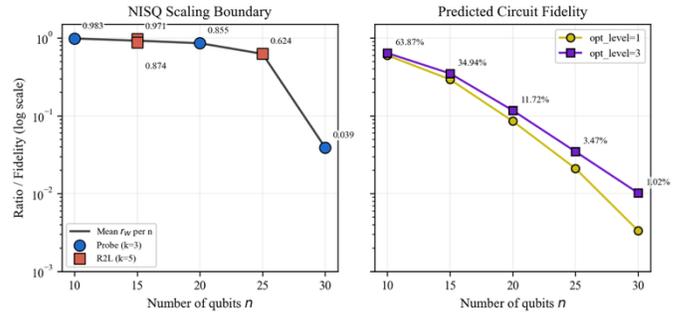

*Figure 6: Comparative scaling of hardware $r_W$ and circuit fidelity on ibm_pittsburgh. Left: Measured WRAcc approximation ratio $r_W$ for Probe (k = 3) and R2L (k = 5) attacks. Right: Predicted circuit fidelity F derived from a layered 2-qubit error rate (EPLG = 2.42 × 10⁻³). The shared log scale highlights that $r_W$ tracks above the fidelity floor until n = 30, where both metrics collapse, marking the operational limit of the NISQ device.*

*4.6 Hybrid IDS Evaluation*

To assess whether the discovered subgroups have operational value beyond interpretability, a two-tier hybrid intrusion detection system is evaluated on the held-out test set. In Tier 1, subgroup conjunction rules are applied as deterministic classifiers: any connection matching a rule is immediately flagged as an attack. Connections not flagged by any rule proceed to Tier 2, where an XGBoost classifier makes the final determination.

On R2L attacks, baseline XGBoost achieves a Detection Rate (DR) = 9.04% (F1 = 0.166); the classical hybrid (30 beam-search rules + XGBoost) reaches DR = 9.32% (F1 = 0.170, Tier-1 precision 99.2%); the quantum hybrid (15 QAOA rules + XGBoost) reaches DR = 12.61% (F1 = 0.222, Tier-1 precision 91.8%). The combined hybrid matches the quantum-hybrid result, indicating that at this qubit scale the quantum rules subsume the classical rules' coverage rather than adding independent signal. The primary value of QAOA-unique subgroups is not aggregate detection-rate improvement, which is marginal at these scales, but the multi-feature interaction patterns they expose for analyst interpretation.

V. DISCUSSION

*5.1 What This Work Demonstrates*

This work establishes the first formulation of subgroup discovery as a QUBO, encoding WRAcc as a quadratic objective ($R^2 = 0.989$) with an Ising Hamiltonian whose ZZ structure produces non-trivial entangling circuits on superconducting hardware. This formulation is novel: all prior quantum work in cybersecurity targets classification accuracy, not combinatorial optimization over interpretable feature subsets. It also provides a dual approximation-ratio framework ($r_E$ for Hamiltonian quality, $r_W$ for application-level WRAcc recovery) for evaluating QAOA on applied problems where the true optimum is known. The empirical scaling characterization on ibm_pittsburgh quantifies the NISQ noise boundary for dense QUBO instances with measured data rather than projections.

The pipeline does not yet outperform classical methods at scales where both run, and we do not claim quantum advantage in the computational sense. At 10 qubits, the noiseless simulator recovers the exhaustive optimum ($r_W = 1.0$) and hardware reaches $r_W = 0.983$, but classical exhaustive search finishes in seconds. At 20 qubits (Probe, $k = 3$), QAOA on hardware reaches $r_W = 0.855$ against a classical baseline that completes in milliseconds. Performance degrades to $r_W = 0.624$ at 25 qubits and $r_W = 0.039$ at 30 as hardware noise dominates. These numbers are not the headline result; they are the boundary against which our other contributions are calibrated.

### 5.2 The QAOA-Unique Subgroup Argument

The strongest empirical result is the discovery of multi-feature R2L attack patterns that beam search prunes. Classical greedy search extends subgroups one feature at a time, retaining only the top WRAcc candidates at each depth, efficient, but blind to patterns where intermediate subgroups score poorly. The six-feature QAOA-unique subgroup combining FTP service indicators, connection-count statistics, and host-behavior diversity metrics achieves 99.6% test precision on R2L attacks: a rule a cybersecurity analyst could immediately enact.

The value at this time is search completeness, not speed. Beam search completes in milliseconds and QAOA takes minutes to hours including cloud queue times, so this is not a runtime argument. It is a coverage argument: at scales where exhaustive enumeration becomes intractable ($n > 50, k > 8$), beam search's miss rate will continue to grow, while QAOA's superposition-based search covers the full combinatorial space in principle. Whether hardware fidelity improves fast enough to realize this in practice is the open question.

### 5.3 Limitations and Honest Assessment

Hardware noise sets a hard ceiling on this pipeline at roughly 25-30 qubits for dense QUBO instances with $O(n^2)$ ZZ terms; current superconducting devices cannot maintain coherence through the required circuit depth. End-to-end runtime is the other practical constraint: cloud queue times, transpilation overhead, and the separate-jobs architecture (a fresh EstimatorV2 instance per COBYLA iteration, used to avoid Session billing) put QAOA at minutes to hours where classical methods finish in seconds to minutes. Both constraints are hardware-side, not algorithmic, and both move with the next generation of devices.

The QUBO surrogate captures 98.9% of WRAcc variance ($R^2 = 0.989$); the residual 1.1% can in principle bias subgroup rankings, and surrogate sampling for $n > 15$ fits a subset of the bitstring space rather than the full landscape. NSL-KDD remains the standard benchmark for IDS work but is dated; modern monitoring produces datasets with hundreds of features and millions of records, beyond current NISQ capabilities. Reaching that level requires either continued hardware progress or feature pre-selection that lowers the qubit requirement without discarding interaction patterns.

## VI. CONCLUSION

### 6.1 Summary of Contributions

This paper presented the first quantum-enhanced subgroup discovery pipeline, formulating the identification of interpretable conjunctive rules as a QUBO solved by QAOA on IBM Quantum hardware. The key contributions are:

(1) **A novel QUBO formulation for subgroup discovery** via least-squares fit to the WRAcc landscape ($R^2 = 0.989$), producing Ising Hamiltonians whose ZZ structure yields non-trivial entangling circuits on superconducting hardware.

(2) **Empirical NISQ scaling characterization on ibm_pittsburgh** at p = 1: $r_W = 0.983$ at 10 qubits, 0.855 at 20, and 0.039 at 30, establishing a measured fidelity boundary for dense QUBO instances.

(3) **Discovery of QAOA-unique multi-feature R2L attack patterns**, combining FTP service indicators, connection-count statistics, and host-behavior diversity metrics, that achieve 99.6% test precision while being systematically pruned by classical beam search.

(4) **A dual approximation-ratio framework** ($r_E$ for Hamiltonian quality, $r_W$ for application-level WRAcc recovery) providing a reusable methodology for evaluating variational quantum algorithms on combinatorial optimization problems.

### 6.2 Future Directions

Three directions would strengthen the case for quantum-enhanced subgroup discovery. First, advanced error mitigation techniques, probabilistic error cancellation, zero-noise extrapolation, or tensor-network error mitigation may extend the useful qubit range beyond the current 20-25 qubit boundary for dense QUBO instances at p = 1. Second, sparse QUBO formulations that reduce the number of ZZ terms per qubit (e.g., by exploiting feature correlation structure to prune weak pairwise interactions) would reduce circuit depth and improve hardware fidelity at larger n. Third, benchmarking on modern IDS datasets (CICIDS2017, UNSW-NB15) with higher feature counts would test the pipeline's scaling behavior in a more realistic setting and determine whether the QAOA-unique subgroups are meaningful at larger problem sizes.

ACKNOWLEDGMENT

This work is supported by the NSF CyberCorps Scholarship for Service (SFS) program under DEG-1946619, and the University of Missouri Quantum Innovation Center (QIC) which provided access to IBM Quantum hardware. The authors acknowledge the use of generative AI tools, specifically Anthropic Claude [30] and Google Gemini [31], for code debugging and refinement during algorithm implementation. All code and results have been independently verified by the authors, who remain fully responsible for the content of this manuscript.